\documentclass[%
 aip,
 jmp,%
 amsmath,amssymb,
preprint,%
]{revtex4-2}

\usepackage{graphicx}
\usepackage{xcolor}
\usepackage{dcolumn}
\usepackage{bm}
\usepackage[normalem]{ulem}
\begin{document}

\preprint{AIP/123-QED}

\title[High-order TSLB]{High-order thread-safe lattice Boltzmann model for HPC turbulent flow simulations}
\author{Andrea Montessori}
\email{andrea.montessori@uniroma3.it}
\author{Michele La Rocca}
 \affiliation{ 
Department of Civil, Computer Science and Aeronautical Technologies Engineering 
} 
\author{Giorgio Amati}
\affiliation{SCAI, SuperComputing Applications and Innovation Department, CINECA, Via dei Tizii, 6, Rome 00185, Italy
} 
\author{Marco Lauricella}
\affiliation{Istituto per le Applicazioni del Calcolo CNR, via dei Taurini 19, 00185 Rome, Italy}
\author{Adriano Tiribocchi}
\affiliation{Istituto per le Applicazioni del Calcolo CNR, via dei Taurini 19, 00185 Rome, Italy}
\affiliation{INFN "Tor Vergata" Via della Ricerca Scientifica 1, 00133 Roma, Italy
} 
\author{Sauro Succi}
\affiliation{Center for Life Nano Science@La Sapienza, Istituto Italiano di Tecnologia, 00161 Roma, Italy
} 
\affiliation{Department of Physics, Harvard University, Cambridge, MA, 02138, USA}


\date{\today}

\begin{abstract}

We present a highly-optimized thread-safe lattice Boltzmann model in which the non-equilibrium part of the distribution function is locally reconstructed via recursivity of Hermite polynomials. Such a procedure allows the explicit incorporation of non-equilibrium moments of the distribution up to the order supported by the lattice. Thus, the proposed approach increases accuracy and stability at low viscosities without compromising performances and amenability to parallelization with respect to standard lattice Boltzmann models.
The high-order thread-safe LB is tested on two types of turbulent flows, namely the turbulent channel flow at $Re_{\tau}=180$ and the axisymmetric turbulent jet at $Re = 7000$, it delivers results  in excellent agreement with reference data (both DNS, theory, and experiments) and a) achieves peak performances ($\sim 5 \; TeraFlop/s$ and an arithmetic intensity of $\sim 7\; FLOP/byte$ on single GPU) by significantly reducing the memory footprint, b) retains the algorithmic simplicity of standard lattice Boltzmann computing and c) allows to perform stable simulations at vanishingly low viscosities.
Our findings open attractive prospects for high-performance simulations of realistic turbulent flows on GPU-based architectures. Such expectations are confirmed by the excellent agreement among lattice Boltzmann, experimental, and DNS reference data.

\end{abstract}


\maketitle

\section{\label{sec:intro} Introduction}

During the last couple of decades, the lattice Boltzmann method (LB hereafter) has met with massive success in the computational fluid dynamics community as a hydrodynamic solver in kinetic disguise due to its strongly appealing computational features combined with its conceptual and practical simplicity \cite{succi,montessori2018lattice}. Indeed, at variance with standard hydrodynamic solvers in which the convective term $u_{\beta}\partial_{\beta}u_{\alpha}$ (where $u_{\alpha}$ is the fluid velocity, and Greek subscripts indicate Cartesian components) is non-linear and non-local at a time, the LB splits non-linearity and non-locality between the two backbones operators, namely the collision (a local and non-linear operator) and the streaming step (linear and non-local and exact at machine precision). In particular, the possibility to advect the set of discrete particle distribution functions along linear characteristics is a fundamental feature when dealing with flows in which the velocity streamlines may become "wild," as it occurs in chaotic, turbulent flows \cite{sreenivasan1999fluid,sreenivasan1989frontiers}.
Nonetheless, the lattice Boltzmann is known to suffer from drawbacks related to different aspects, where two of the major ones are the following:

1) From the computational standpoint, the LB possesses a relatively low operational (or arithmetic) intensity, in the range $1 \to 4$ FLOP/byte (floating point operations needed to compute a kernel per byte of memory accessed), depending on the collision strategy and the floating point representation employed \cite{exasc2}. Such a low operational intensity prevents exploiting the ideal peak performance that processors and graphical accelerators could potentially deliver. 

2) The LB model, in particular the widely used version based on the BGK single relaxation collisional operator \cite{BGK}, is known to suffer from the onset of instabilities at low viscosity (over-relaxed regimes) due to the relaxation of non-hydrodynamic ghost moments \cite{montessori2014regularized} which, in turn, stem from the lack of sufficient isotropy and moment representation of low order lattices.
Alternative formulations based on more complex collisional operators have been proposed \cite{d2002multiple, geier2006cascaded} to deal with such shortcomings, albeit at the expense of simplicity and performances of the single relaxation time LB. These drawbacks limit the use of BGK-like LB models to the simulation of low Reynolds number values and creeping flows since their capability to deal with turbulent flows can only be guaranteed upon increasing the grid resolution.

In this work, we propose an LB model built upon a recently developed thread-safe version \cite{MONTESSORIjocs23} and a higher order variant of the regularized LB \cite{malaspinas2015increasing}, which is aimed at optimizing the performances on shared memory architecture typical of graphic processing units and to broaden the range of applicability of BGK-based LB to low viscosity fluid flows.
The high-order thread-safe LB is tested on two types of turbulent flows, namely the turbulent channel flow at $Re_\tau=180$ and the axisymmetric turbulent jet at $Re=7000$. The results are in excellent agreement with reference data (both DNS, theory, and experiments) and confirm the capability of the approach to perform accurate high-performance computer simulations of turbulent flows of scientific and engineering interest.

\section{\label{sec:Method} Method}

In this section, an in-depth description of the proposed computational framework based on the cross-fertilization between a recently developed thread-safe LB strategy \cite{MONTESSORIjocs23} and a high-order variant of the regularized LB exploiting the recursivity of Hermite polynomials \cite{grad1949kinetic,grad1949note, malaspinas2015increasing} is provided.
Before proceeding, we shortly recap the salient features of the lattice Boltzmann equation in the single-relaxation time approximation (a thorough review of the LB equation can be found at Ref. \cite{kruger2017lattice, succi2018, montessori2018lattice}).

We start from the following discrete version of the Boltzmann equation
\begin{equation} \label{lbe}
    f_i(x_\alpha+c_{i\alpha}\Delta t,t + \Delta t)=f_i(x_\alpha,t) + \omega(f^{eq}_i(\rho,\rho u_\alpha)-f_i(x_\alpha,t)),
\end{equation}
where $x_\alpha$ and $t$ denote lattice position and time step, respectively, $f_i(x_\alpha,t)$ is a set of distribution functions, $i=1,..,q$ is an index spanning the $q$ discrete velocity vectors $c_{i\alpha}$ of the lattice and $f_i^{eq}(\rho,\rho u_\alpha)$ is a set of discrete thermodynamic equilibria obtained via a second-order expansion in Mach number of the continuous Maxwell-Boltzmann distribution. Also, $\rho$ is the fluid density, and $\rho u_{\alpha}$ is the fluid momentum. 
An analytical, truncated expression of $f_i^{eq}$ is given by
\begin{equation}\label{lbequil}
    f_i^{eq}=w_i \rho\left(1 + \frac{c_{i\alpha} \cdot u_\alpha}{c_s^2} + \frac{(c_{i\alpha} c_{i\beta} - c_s^2\delta_{\alpha\beta})u_\alpha u_\beta}{2c_s^4}   \right ),
\end{equation}
where $w_i$ is a normalized set of weights and $c_s^2=1/3$ is the sound speed of the model. The standard LB algorithm consists of two main steps, namely a collision towards a local Maxwell-Boltzmann equilibrium occurring at a characteristic rate $\omega$ (right-hand side of Eq.(\ref{lbe})) and a free-flight of the set of distributions along predefined lattice directions. The first operation is local and non-linear in the macroscopic velocity, while the second one is a linear non-local operation, exact at machine precision (no loss of information during advection). 
The main hydrodynamic fields of interest, i.e., density $\rho$, linear momentum $\rho u_\alpha$ and momentum flux tensor $\Pi_{\alpha\beta}$, can be calculated from the kinetic moments of the distribution functions $f_i$ by linear and local summations \cite{succi2018}. 
Finally, through a multi-scale expansion in the Knudsen number \cite{chapman1990mathematical}, it is possible to prove that the lattice Boltzmann equation recovers, in the continuum limit, a set of partial differential equations governing the conservation of mass and momentum of a volume of fluid, in the limit of weak compressibility (small Mach numbers)
\begin{equation}
    \partial_t \rho +  \partial_{\alpha}\rho u_\alpha=0,
\end{equation}
\begin{equation}
    \frac{\partial \rho u_\alpha}{\partial t} + \partial_{\beta} (\rho u_\alpha u_\beta) = \partial_{\alpha}(-p \delta_{\alpha\beta} + \rho\nu(\partial_{\beta} u_\alpha + \partial_{\alpha} u_\beta)),
\end{equation}
where $p=\rho c_s^2$ is the pressure, and $\nu=c_s^2(1/\omega-0.5)$ is the kinematic viscosity. 
\subsection{\label{sec:threadsafe} Thread-safe reconstruction of post-collision  distributions }

The main idea behind the thread-safe (fused streaming-collision) LB algorithm comes from observing that, in a lattice BGK relaxation process, any post-collision distribution can be expressed as a weighted sum of the equilibrium and non-equilibrium parts of the $i^{th}$ probability distribution function (pdf), as follows \cite{succi}:
\begin{equation}\label{LBexp}
    f^{post}_{i} =f^{pre}_{i} + \omega(f^{eq}_{i} - f^{pre}_{i})=f^{eq}_{i} + (1-\omega)f^{neq}_{i}.
\end{equation}
The above equality stems from the decomposition of the full lattice distribution into an equilibrium and non-equilibrium part, as $f_i=f_i^{eq} + f_i^{neq}$. 

A direct consequence of Eq.(\ref{LBexp}) is the possibility of reconstructing the set of distribution functions by decoupling pre and post-collisional states, with evident advantages for the implementation of highly efficient LB models on shared-memory architectures (SMA) present in graphic processing units (GPUs), as previously shown in \cite{mattila, latt2021cross}.
Indeed, Eq.(\ref{LBexp}) can be easily incorporated within actual lattice Boltzmann codes to eliminate data dependencies that arise during non-local read and write operations. Such data dependencies hinder an efficient  implementation of thread-safe LB models on SMA unless more complex streaming strategies \cite{bailey2009accelerating, geier2017esoteric, lehmann2022esoteric}, 
which inevitably compromise the underlying simplicity of the LB, are employed. 


Thus, a natural extension of Eq.(\ref{LBexp}) calls for an explicit inclusion of the streaming step as follows:

\begin{equation}\label{pushLB}
    f_{i}(x_\alpha+c_{i\alpha},t+1) = f^{eq}_{i}(x_\alpha,t) + (1-\omega)f^{neq}_{i}(x_\alpha,t).
\end{equation}

It is worth noting that Eq.(\ref{pushLB}) is a thread-safe operation on SMA provided that $f_i^{neq}$  is directly calculated from macroscopic hydrodynamic quantities which, as noted above, in the LB framework can be efficiently computed as statistical moments of the set of PDFs, up to an order dictated by the symmetry and isotropy properties of the lattice. 
Such a task can be performed by resorting to the regularization procedure\cite{montessori2014regularized, montessori2015lattice, montessori2018regularized, latt2006, zhang2006efficient}, where the set $f_i^{neq}$ is defined as a 
projection of the hydrodynamic moments $\rho$, $\rho u_\alpha$ and $\Pi^{neq}_{\alpha\beta}\propto(\partial_{\beta} u_\alpha + \partial_{\alpha} u_\beta)$ onto suitable Hermite basis.

More precisely, using Hermite polynomials and Gauss-Hermite quadratures, one can write $f^{neq}_i=w_i \sum_n \frac{1}{c_s^{2n} n!}a_{\alpha 1...\alpha n}^n \mathcal{H}_{i,\alpha1...\alpha n}^n$ where $\mathcal{H}^n$ is the n-th order Hermite basis and $a_{\alpha1...\alpha n}^n=\sum_i f^{neq}_i \mathcal{H}_{i,\alpha1 ... \alpha n}^n$ the corresponding Hermite expansion coefficients. Both $a^n$ and  $\mathcal{H}^n$ are rank $n$ tensors.
In standard regularized LB models, the non-equilibrium set of distributions contains information from hydrodynamic moments up to order 2 (i.e. the Navier-Stokes level) and is filtered from non-hydrodynamic (ghost) contributions stemming from higher order fluxes \cite{dellar2006non, zhang2006efficient}. 
Thus, the set of non-equilibrium distributions can be compactly written as:

\begin{equation}
    f^{neq}_i=\frac{w_i}{2 c_s^4} (c_{i\alpha}c_{i\beta} - c_s^2\delta_{\alpha\beta}) \Pi^{neq}_{\alpha\beta}
\end{equation}

Since, at the continuum level, the non-equilibrium pdfs $f_i^{neq}$ contain information of non-equilibrium moments at all orders, they can be re-projected onto suitable Hermite basis to include a larger number of hydrodynamic moments, up to the order of isotropy of the lattice.

With the above relation in mind, the lattice kinetics exposed in Eq.(\ref{LBexp}) can be implemented by storing, at each time step and for each lattice node, three macroscopic quantities, namely, one scalar ($\rho$), three vector components ($\rho \mathbf{u}$) and six components of a symmetric, tensor of rank two ($\Pi^{neq}_{\alpha\beta}$). 
Thus, the full stream and collision step becomes

\begin{multline} \label{fullstrcoll}
    f_{i}(x_\alpha+c_{i\alpha},t+1) = \rho w_i [ 1 + \frac{c_{i\alpha}\cdot u_\alpha}{c_s^2} + \frac{(c_{i\alpha}c_{i\beta}-c_s^2\delta_{\alpha\beta}) u_\alpha u_\beta}{2 c_s^4}] + \\ (1 - \omega)\frac{w_i}{2 c_s^4} (c_{i\alpha}c_{i\beta} - c_s^2\delta_{\alpha\beta}) \Pi^{neq}_{\alpha\beta}.
\end{multline}

\begin{figure}
    \centering
    \includegraphics[scale=0.8]{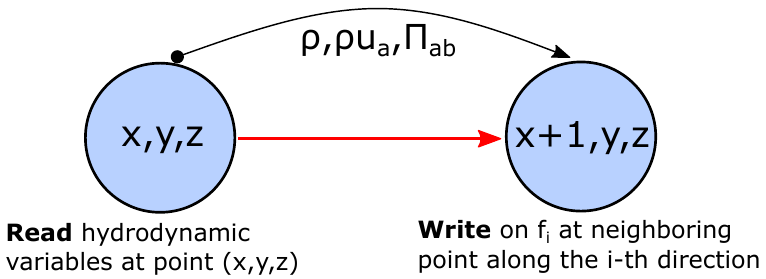}
    \caption{Sketch of the thread-safe LB update algorithm. Rather than explicitly streaming distributions along the lattice directions, the algorithm reconstructs the post-streamed distribution using the neighboring macroscopic hydrodynamic fields, which are employed to rebuild the full set of distributions.}
    \label{fig:threadsafe}
\end{figure}

In other words, the thread-safe LB, rather than explicitly streaming distributions along the lattice directions, aims at reconstructing the post-streamed/post-collided distribution via a scattered reading of the neighboring macroscopic hydrodynamic fields, which are then employed to rebuild the complete set of distributions (see Fig.\ref{fig:threadsafe}). 
It is interesting to note that Eq.(\ref{fullstrcoll}) clearly 
allows for the implementation of the lattice Boltzmann 
on unstructured grids and meshless domains due to the possibility of interpolating macroscopic fields (rather than the full set of PDFs) to reconstruct the distributions off-lattice. Such extension is currently ongoing and will be the subject of separate works.

As anticipated in the introduction, a fundamental feature of the thread-safe approach lies in its simplicity with respect to other existing single-distribution, in-place streaming approaches, such as EsoTwist \cite{geier2017esoteric, lehmann2022esoteric} and the Bailey algorithm \cite{bailey2009accelerating}, which require the resort to high-level programming due to the need of explicitly swapping pointers to preserve data locality.
Indeed, since the post-collision distributions lose any connection with their pre-collision values, the streaming step is replaced by a scattered reconstruction of the post-collisionals, performed at neighboring points along the $i^{th}$  lattice direction upstream.
From a computational standpoint, such an approach is particularly beneficial, as it avoids the onset of race conditions that may jeopardize the correct memory access in shared memory GPU architectures \cite{MONTESSORIjocs23}. 
\begin{figure}
    \centering
    \includegraphics[scale=0.5]{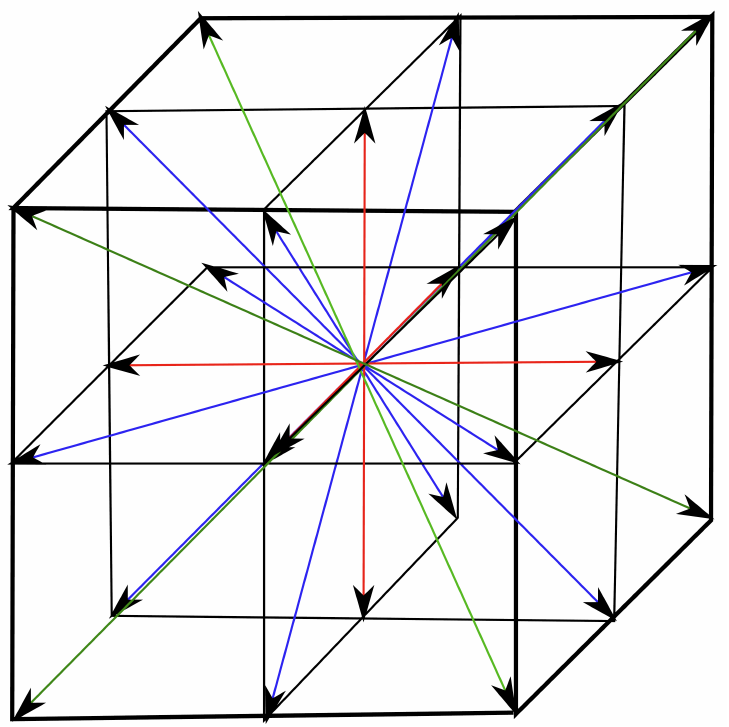}
    \caption{Graphical representation of the D3Q27 lattice. Red arrows stand for the velocity group having magnitude $|c|=\sqrt{c_x^2+c_y^2+c_z^2}=1$, the blue one $|c|=\sqrt{2}$ and the green arrows $|c|=\sqrt{3}$ }
    \label{fig:d3q27}
\end{figure}
Moreover, the thread-safe reconstruction allows for a straightforward porting of LB codes on GPU, either by resorting to CUDA native language or by employing directive-based procedures, in the second case, with virtually no need to modify the structure of the serial code. The latter strategy has been employed in this work.

We highlight that, in separate contexts, the use of regularized post-collisionals was proven to be instrumental in accessing off-equilibrium hydrodynamic regimes simply out of reach for standard, second-order isotropic lattice Boltzmann models and increasing the stability and accuracy of LB models to high inertia-over-dissipation hydrodynamic regimes,  typical of turbulent flows \cite{montessori2015lattice, montessori2018regularized, malaspinas2015increasing, montessori2019jfm}. 


In the next subsection, we present a thread-safe framework in which third-order discrete equilibria are included where the set of non-equilibrium distributions is expanded to 
account for non-equilibrium statistical moments up to order three. This is the highest moment isotropy order supported by the standard fourth-order isotropic lattices, i.e., the $D3Q27$ sketched in figure \ref{fig:d3q27}. 


\subsection{\label{sec:recreg} High-order regularized lattice Boltzmann based on recursivity of Hermite polynomials}

The Hermite relation $ f^{neq}_i=w_i \sum_n \frac{1}{c_s^{2n} n!}a_{\alpha 1...\alpha n}^n \mathcal{H}_{i\alpha1...\alpha n}^n$ represents a hierarchy of terms with which the non-equilibrium part of the distribution functions can be thoroughly described, provided that the lattice in use supports an exact moment representation up to order $n$ \cite{chen2008discrete}. 

As discussed in the previous section, the distribution can be split into its equilibrium and non-equilibrium parts, which can be represented through Hermite polynomials:

\begin{equation}
    f_i^{eq}=w_i \sum_n \frac{1}{c_s^{2n} n!}a_{eq,\alpha 1...\alpha n}^n \mathcal{H}_{i\alpha1...\alpha n}^n
\end{equation}
\begin{equation}
    f_i^{neq}=w_i \sum_n \frac{1}{c_s^{2n} n!}a_{neq,\alpha 1...\alpha n}^n \mathcal{H}_{i\alpha1...\alpha n}^n.
\end{equation}

The equilibrium and non-equilibrium Hermite coefficients are defined as:

\begin{equation} \label{hermite_eq}
   {a}_{eq,\alpha 1,...,\alpha n}^n=\sum_i \mathcal{H}_{i\alpha 1,...,\alpha n}^n f_i^{eq},
\end{equation}
and 
\begin{equation}
    a_{neq,\alpha 1,...,\alpha n}^n=\sum_i \mathcal{H}_{i\alpha 1,...,\alpha n}^n f_i^{neq}.
\end{equation}

As far as the equilibrium Hermite coefficients are concerned, one can construct the $n-th$ order coefficient by exploiting the following relation \cite{shan2006kinetic}:

\begin{equation}
    {a}_{eq,\alpha 1,...,\alpha n}^n={a}_{eq,\alpha 1,...,\alpha n-1}^{n-1}u_{\alpha n},
\end{equation}

being ${a}_{eq}^0=\rho$, $a_{eq}^1=\rho u_\alpha$ and $a_{eq}^2=\rho u_\alpha u_\beta + \rho c_s^2 \delta_{\alpha\beta}$ .
The D3Q27 (and the D2Q9 counterpart in 2D) correctly recovers macroscopic moments up to order three ($n=3$)\cite{succi, kruger2017lattice}. Thus, by employing the relation (\ref{hermite_eq}) expanded up to $n=3$ \cite{shan2006kinetic}, higher order hydrodynamic equilibria become

\begin{multline}\label{eqts}
    f_i^{eq}=\rho w_i [ 1 + \frac{c_{i\alpha}\cdot u_\alpha}{c_s^2} + \frac{(c_{i\alpha}c_{i\beta}-c_s^2\delta_{\alpha\beta}) u_\alpha u_\beta}{2 c_s^4} \\ 
    + \frac{(c_{i\alpha}c_{i\beta}c_{i\gamma}-c_{i\gamma}c_s^2\delta_{\alpha\beta} - c_{i\alpha}c_s^2\delta_{\beta\gamma} - c_{i\beta}c_s^2\delta_{\alpha\gamma}) u_\alpha u_\beta u_\gamma}{6 c_s^6}].
\end{multline}

These third-order discrete Maxwell-Boltzmann equilibrium distributions allow, when equipped to a fourth-order isotropic lattice, to recover hydrodynamic moments up to order $3$, i.e.,

\begin{equation}
    Q_{\alpha\beta\beta}=\sum_i f_i^{eq}c_{i\alpha}c_{i\beta}c_{i\beta},
\end{equation}

which can be interpreted as a flux of momentum flux \cite{montessori2015lattice}. It is important to recall that, despite such a higher-order moment having no counterpart in the Navier-Stokes equation, the use of extended equilibria increases both the stability and accuracy of an LB solver.

Moreover, as shown in \cite{grad1949note, shan2006kinetic}, it is also possible to define recursive relations for the non-equilibrium Hermite coefficients, which can be compactly written as:

\begin{equation}
    a^n_{{neq},\alpha 1,...\alpha n}= a^{n-1}_{{neq},\alpha 1,...\alpha n-1} u_{\alpha n} + (u_{\alpha 1} \cdot\cdot\cdot u_{\alpha n-2} a^{2}_{{neq},\alpha n-1 \alpha n}).
\end{equation}

Recalling that $a_{{neq},\alpha\beta}^{2}=-\frac{1}{\omega c_s^2} (\partial_{\alpha}\rho u_\beta+\partial_{\beta}\rho u_\alpha))$, one can explicitly reconstruct the non-equilibrium set of distributions by incorporating non-equilibrium hydrodynamic moments up to the third-order

\begin{multline} \label{noneqts}
    f_i^{neq}=\frac{c_{i\alpha}c_{i\beta} a^2_{1,\alpha\beta}}{2 c_s^4} + \\ \frac{(c_{i\alpha}c_{i\beta}c_{i\gamma}-c_{i\gamma}c_s^2\delta_{\alpha\beta} - c_{i\alpha}c_s^2\delta_{\beta\gamma} - c_{i\beta}c_s^2\delta_{\alpha\gamma}) (a^2_{{neq},\alpha\beta} u_\gamma + u_\alpha a^2_{{neq},\beta\gamma} + u_\beta a^2_{{neq},\alpha\gamma} )}{6 c_s^6},
    \end{multline}
    
and thus finally obtain
the post-collision and streaming distributions via recursive thread-safe LB  
\cite{MONTESSORIjocs23}:

\begin{multline}
    f_i(x_{\alpha}+c_{i\alpha} \Delta t,t+\Delta t)= \rho w_i ( 1 + \frac{c_{i\alpha}\cdot u_\alpha}{c_s^2} + \frac{(c_{i\alpha}c_{i\beta}-c_s^2\delta_{\alpha\beta}) u_\alpha u_\beta}{2 c_s^4} \\ 
    + \frac{(c_{i\alpha}c_{i\beta}c_{i\gamma}-c_{i\gamma}c_s^2\delta_{\alpha\beta} - c_{i\alpha}c_s^2\delta_{\beta\gamma} - c_{i\beta}c_s^2\delta_{\alpha\gamma}) u_\alpha u_\beta u_\gamma}{6 c_s^6}  ) + (1-\omega)(\frac{c_{i\alpha}c_{i\beta} a^2_{1,\alpha\beta}}{2 c_s^4} + \\ \frac{(c_{i\alpha}c_{i\beta}c_{i\gamma}-c_{i\gamma}c_s^2\delta_{\alpha\beta} - c_{i\alpha}c_s^2\delta_{\beta\gamma} - c_{i\beta}c_s^2\delta_{\alpha\gamma}) (a^2_{{neq},\alpha\beta} u_\gamma + u_\alpha a^2_{{neq},\beta\gamma} + u_\beta a^2_{{neq},\alpha\gamma} )}{6 c_s^6}).
\end{multline}


It is worth noting that the cross-fertilization between recursivity and thread-safe strategy permits the implementation of a higher-order LB model by sensibly reducing the memory footprint without altering the simplicity of the LB scheme based on the AB (flip-flop) streaming strategy. Indeed, in an actual implementation of a high-order thread-safe LB, one needs to store just one set of distributions (i.e., one set per fluid component) plus, in three dimensions, 10 arrays for the macroscopic variables, namely, density, linear momentum, and non-equilibrium momentum flux tensor. The higher-order Hermite coefficients can then be computed locally and on the fly while performing the thread-safe fused stream-and-collide algorithm.

\subsection{\label{sec:threadsafeBCs} Thread-safe implementation of the non-equilibrium extrapolation boundary conditions}

In this section, we present a novel strategy to implement boundary conditions based on the concept of non-equilibrium extrapolation \cite{guo2002extrapolation}, together with the possibility to approximate missing data at the boundary through Grad's reconstruction\cite{chikatamarla2006grad}. Such implementation can be naturally embedded in the thread-safe LB framework and employed to model different boundary conditions accurately, such as Dirichlet-like (no-slip inflow, imposed velocity) and Neumann-like (outflow, zero-gradient).

The non-equilibrium extrapolation method is based on the observation that the post-collision distribution at a boundary node can be rewritten as:

\begin{equation} \label{original_neq}
    f_i(x_{B\alpha},t)=f_i^{eq}(x_{B\alpha},t)+(1-\omega)f_i^{neq}(x_{B\alpha},\tau),
\end{equation}

where $x_{B\alpha}$ denotes a boundary node. As in the regularized model, it is possible to assume that the $f_i^{neq}=\epsilon f_i^1$, 
where $\epsilon$ is the Knudsen number.
Since, at time $t$, the hydrodynamic quantities are known at a neighboring fluid node ($x_{F\alpha}$), $f_i^{neq}(x_{F\alpha},t)$ can be determined as:

 \begin{equation}
     f_i^{neq}(x_{F\alpha},t)=f_i(x_{F\alpha},t)-f_i^{eq}(x_{F\alpha},t).
 \end{equation}

By observing that the boundary node $x_{B\alpha}$ is at a distance $\Delta x=\epsilon +c_{i\alpha}\Delta t$ from the nearest bulk node, one can write $f_i^{neq}(x_{B\alpha},t)=f_i^{neq}(x_{F\alpha},t) + \mathcal{O}(\epsilon)$, thus $f_i^{neq}(x_{B\alpha},t)$ can be approximated by performing a first order extrapolation as

 \begin{equation} \label{neq_app}
     f_i^{neq}(x_{B\alpha},t)= f_i(x_{F\alpha},\tau)-f^{eq}_i(x_{F\alpha},\tau) +\mathcal{O}(\epsilon^2).
 \end{equation}
The above implies that the non-equilibrium distribution at the boundary is approximated at second-order accuracy.

Finally, Eq.(\ref{original_neq}) can be rewritten as

\begin{equation} \label{neq_final_guo}
    f_i(x_{B\alpha},t)=f_i^{eq}(x_{B\alpha},t)+(1-\omega)(f_i(x_{F\alpha},t)-f^{eq}_i(x_{F\alpha},t)),
\end{equation}
where the equilibrium distribution is computed by imposing the hydrodynamic variables at the boundary (either momentum or density/pressure), while the non-equilibrium part is extrapolated from the neighboring bulk nodes. 
\begin{figure}
    \centering
    \includegraphics[scale=0.8]{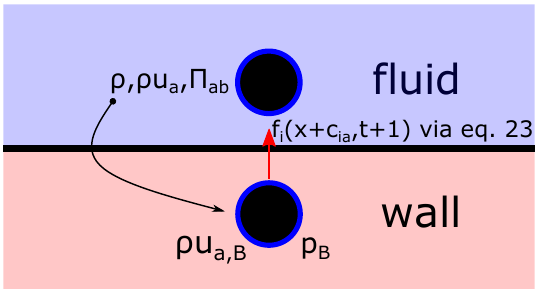}
    \caption{Thread-safe, explicit non-equilibrium reconstruction of the boundary condition. The non-equilibrium part of the incoming (unknown) distributions from the boundary nodes is reconstructed using the algorithm reported in eq.\ref{neq_final}. The equilibrium part is computed by imposing macroscopic variables at the boundary node.}
    \label{fig:neqthread}
\end{figure}
The post-collision distribution at the boundary node is then streamed back within the fluid domain. 

These boundary conditions can be naturally incorporated within the thread-safe LB framework.
Indeed, considering that 
the non-equilibrium distribution of Eq.(\ref{neq_app}) can be replaced by an explicit reconstruction through Hermite projection,
Eq.(\ref{neq_final_guo}) can be rewritten as:
\begin{equation} \label{neq_final}
    f_i(x_{B\alpha} + c_{i\alpha},t+1)=f_i^{eq}(x_{B\alpha},t)+(1-\omega)t_i\sum_n \frac{1}{c_s^{2n} n!}\mathcal{H}_{i\alpha 1...\alpha n}^n a_{1\alpha 1...\alpha n}^n(x_{F\alpha},t)
\end{equation}

Thus, the non-equilibrium part of the (unknown) distributions incoming from the boundary nodes is readily reconstructed by employing the algorithm reported in Eq.(\ref{neq_final}), while the equilibrium part is evaluated from the macroscopic variables defined at the boundary node (see Fig.\ref{fig:neqthread}).

It is important to recall that the standard implementation of Eq.(\ref{original_neq})  requires two sets of distributions to be stored in memory in order for the non-equilibrium PDFs at the boundary to be safely evaluated on SMA. In addition, the explicit reconstruction of $f^{neq}$ reported above can be naturally incorporated within the proposed computational framework, thus allowing the use of a data structure compliant with the thread-safe strategy and halving the memory footprint with respect to the standard fused AB scheme.
Finally, the modified non-equilibrium
boundary condition, whose calculation is based on the explicit evaluation of the non-equilibrium distributions from Hermite polynomials, naturally filters out the ghost moments (transported by the streaming step in a BGK LB), which are known to jeopardize the stability and accuracy of LB models operating in both under and over-relaxed regimes.
For the reason above, the modified boundary condition is expected to perform well 
in turbulent flow applications, where LB  generally 
operates at vanishing-low viscosities.

Another important advantage of the present 
approach lies in the possibility of imposing boundary conditions straightforwardly on 
pressure and momentum, as in standard Navier-Stokes solvers, while approximating the momentum flux tensor (containing information on the deviatoric part of the stress tensor) with its value in the nearest-neighbor bulk node. 
This essentially means that the boundary conditions for the velocity and/or pressure do not need to be translated into conditions for the density distribution
functions, a procedure that can be cumbersome for problems involving complex boundaries.
A full assessment of the computational features of the novel thread-safe non-equilibrium boundary conditions is currently ongoing and will be the subject of a separate publication.

\section{\label{sec:Results} Benchmarks}

In this section, we test the accuracy of the high-order thread-safe LB against two types of turbulent flows, namely the turbulent flow in a straight, smooth channel at $Re_{\tau}=180$ and the axisymmetric turbulent jet at $Re_{jet}=7000$. All the simulations have been run in single precision on the Leonardo cluster (single GPU). 

\subsection{\label{sec:channel} Turbulent channel flow at $Re_{\tau}=180$ }

The first test case, referred to as turbulent channel flow, is the flow between two parallel smooth walls driven by a constant upstream-downstream pressure gradient
\cite{marusic2010wall}, with periodic boundary conditions along the directions parallel to the walls (i.e., along the $x$ and $y$ directions).
Due to the periodicity of the momentum along the streamwise direction (i.e., $x$ coordinate), the pressure gradient driving the flow can be mimicked by imposing a constant body force (i.e., gravity) acting along the same direction. Also, no-slip boundary conditions are enforced at the top and bottom walls (i.e., at $z=H$ and $z=-H$, H standing for the half-width of the channel, see top panel of Fig.\ref{fig:loglaws}). At sufficiently high Reynolds numbers, the transition to turbulence is observed, and the flow becomes fully three-dimensional and chaotic. 
The results investigated hereafter refer to a regime of statistically steady turbulence, where the turbulent quantities, obtained as first and second order statistics of the hydrodynamic quantities (i.e. the three components of the velocity field $u_\alpha$, $\alpha=x,y,z$), reach the steady state in a statistical sense.

Lattice Boltzmann results have been compared to reference data obtained from a Chebyshev pseudo-spectral method \cite{kim1987turbulence} applied to the direct numerical simulations of the turbulent channel flow at friction Reynolds number $Re_\tau=H u_\tau/ \nu=180$, where $u_\tau$ is the friction velocity defined as $u_\tau=\sqrt{\tau_w/\rho}=\sqrt{g H}$. Here $g$ is the constant acceleration imposed on the flow and $\tau_w$ the wall friction. Such value of the friction Reynolds number is high enough to guarantee a developed turbulent regime within the channel. 

The simulations have been run on a computational domain $\alpha_x H \times \alpha_y H \times 2H $ where $\alpha_x=24$ and $\alpha_y=1$. Two different resolutions have been employed, namely $H=64$ and $H=128$ lattice units, delivering values of the grid step size in wall units  $\Delta^+=3.8$ and $\Delta^+=1.4$ (being $\Delta^+=Re_\tau/H$). In particular, as reported in \cite{pope2000turbulent} for the case of a fully developed channel flow, a value of $\Delta^+<1.5$ should guarantee a fully resolved simulation. Nonetheless, the thread-safe LB has allowed us to perform stable simulations at resolutions as low as $\Delta^+\sim 6$ (not reported in this work).
The value of the acceleration $g$ imposed to the fluid has been fixed to $10^{-7}$ ($(lattice \, units(lu))/step^2$), while the viscosity can be inferred from the $Re_\tau$ and is $\nu=0.0009\;lu^2/step$ for $H=64$ and $\nu=0.0018$ for $H=128$ (hereafter, all the dimensions are reported in lattice units \cite{latt2008choice}).
The force has been implemented 
as in \cite{luo2000theory}, through the local forcing term $\frac{ \rho w_i c_{i \alpha} F_\alpha}{c_s^2}$ directly added during the thread-safe stream-collision step.

\subsubsection{Mean velocity, Reynolds stresses and Skewness}\label{Mean velocity, Reynolds stresses and Skewness}

\begin{figure}
    \centering
    \includegraphics[scale=0.65]{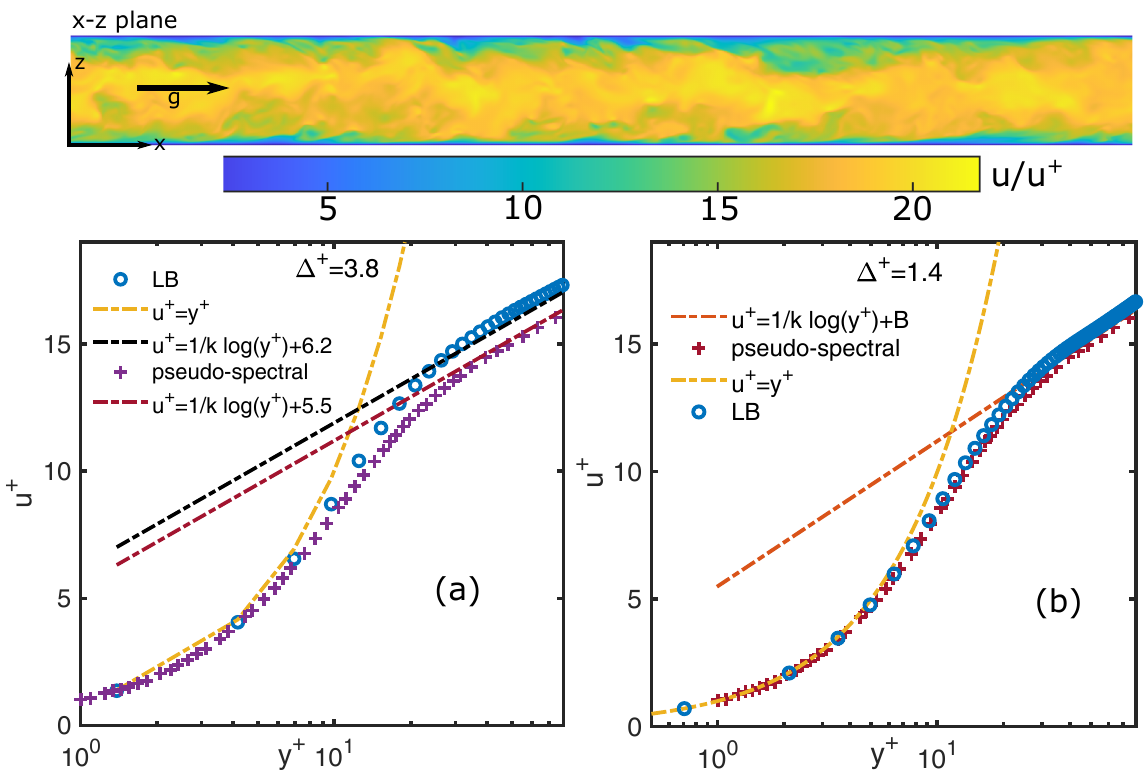}
    \caption{Top panel: Instantaneous velocity field of the turbulent channel flow at $Re_\tau=180$ (the color bar denotes the velocity values in wall units). Bottom panels: The plots show the space-time averaged velocity profiles (made non-dimensional by the friction velocity $u_\tau$) for two different grid resolutions (left plot $\Delta^+=3.8$, $H=64$ grid nodes, right plot $\Delta^+=1.4$, $H=128$). The legend of the symbols is shown in the plots.}
    \label{fig:loglaws}
\end{figure}

The first statistic investigated is the mean velocity profile. In Fig.\ref{fig:loglaws} we report the mean streamwise velocity profiles (scaled by the friction velocity $u_{\tau}=\sqrt{gH}$) obtained for two different resolutions $\Delta^+=3.8$ (Fig.\ref{fig:loglaws}a, $2H=128$) and $\Delta^+=1.4$ (Fig.\ref{fig:loglaws}b, $2H=256$). 
The LB results are compared with the DNS pseudo-spectral results reported in \cite{kim1987turbulence} (plusses), the near-wall linear solution $u^+=y^+$ and the logarithmic law in bulk $u^+=\frac{1}{k}log(y^+) + B$ (dashed lines), being the Von Karman constant $k\sim0.41$ and $B\sim 5 $ for smooth walls.

\begin{figure}
    \centering
    \includegraphics[scale=0.7]{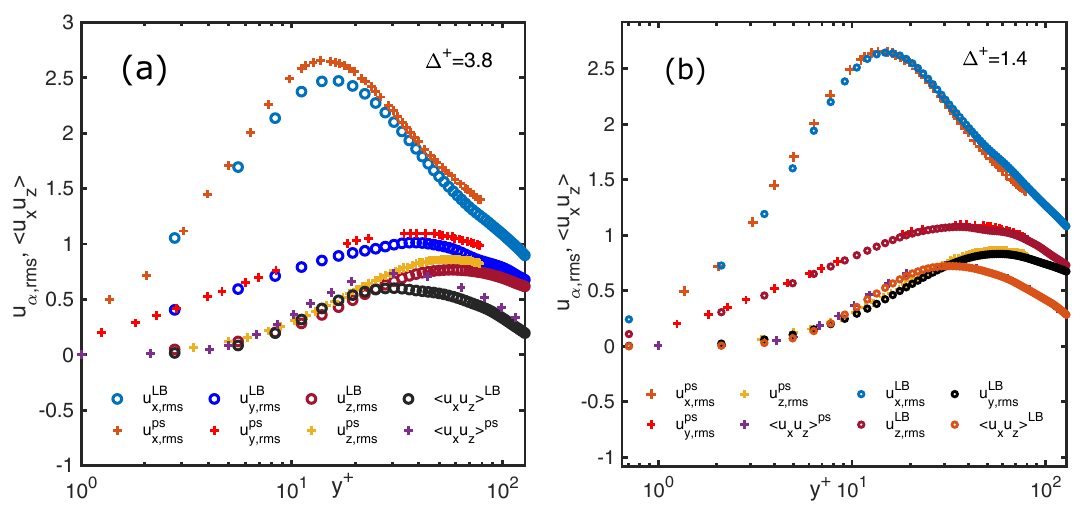}
    \caption{Mean turbulent kinetic energy components and Reynolds stresses. LB data are compared against the pseudo-spectral results reported in \cite{kim1987turbulence}. In panel (a) $\Delta^+=3.8$, in panel (b) $\Delta^+=1.4$.}
    \label{fig:stresses}
\end{figure}

The near-wall linear behavior is correctly reproduced at both resolutions. In contrast, the simulation at $\Delta^+=3.8$ slightly overestimates the DNS solution in bulk, as can be observed from Fig.\ref{fig:loglaws}a, being it described by a log-law with $B\sim 6$, a value larger than the one predicted for turbulent channel flows in the presence of smooth walls. As expected, by increasing the resolution, DNS data can be reproduced accurately, as shown in Fig.\ref{fig:loglaws}b. This is not surprising since, as reported in \cite{pope2000turbulent}, a uniform grid step size $\Delta^+<=1.5$ is required for fully resolved simulations, while $\Delta^+<2.5$ is a safe value that guarantees stable simulations. Nonetheless, as far as the mean velocity profile is concerned, the thread-safe high-order LB keeps on providing quite satisfactory results even at coarser resolutions.

We now focus on the second-order (turbulent kinetic energy components and Reynolds stresses) and third-order (skewness) velocity statistics. The first one is defined as:

\begin{equation}
    S_{2,ij}=\frac{<u'_{i}u'_j>}{u_\tau^2},
\end{equation}

where the diagonal components ($i=j$) stand for the root mean square values of the turbulent kinetic energy components, while the off-diagonal terms are the Reynolds stresses.

The second one, i.e., the skewness of individual velocity components, is defined as follows:

\begin{equation}
    S_{3,i}=\frac{<u_{i}^{'3}>}{<u_i^{'2}>^{3/2}}.
\end{equation}

In order to obtain converged statistics, the simulations have been run long enough (several millions of cycles) to reach a statistically steady state.

In Fig.\ref{fig:stresses}, the agreement between the reference and the LB results is noticeable, even for the coarser simulation (panel (a)), where a larger discrepancy appears near the peak of the r.m.s x-component of the turbulent kinetic energy. Nonetheless, the thread-safe LB 
provides 
reliable results despite the discretization in use ruling out the possibility of capturing the turbulent structures in full throughout all the relevant scales. Such a discrepancy is, as expected, much reduced in the simulation performed at $\Delta^+=1.4$ as clearly visible in panel (b), thus confirming that the TSLB simulation in the DNS regime allows to correctly reproduce the  salient 
features 
of the turbulent channel flow, in agreement with reference data.

\begin{figure}
    \centering
    \includegraphics[scale=0.55]{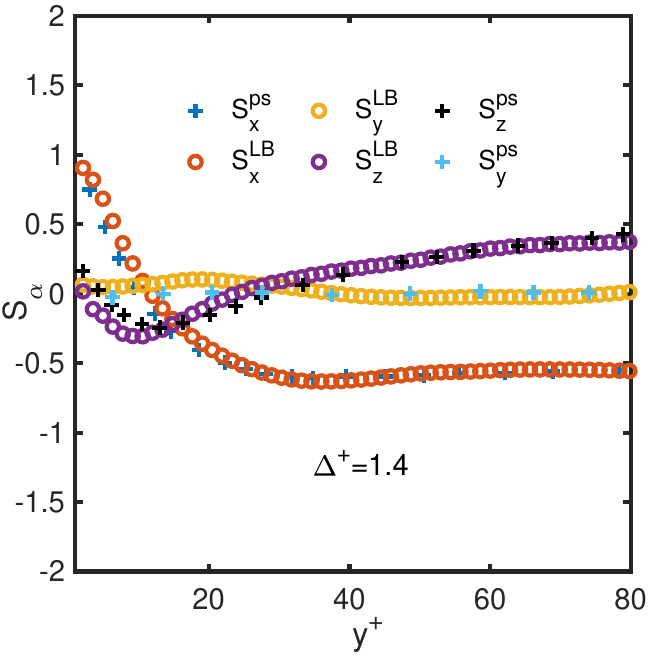}
    \caption{The plot shows the values of the three components of the skewness at $\Delta^+=1.4$ of the fluctuating velocity distributions. LB data are compared against the reference pseudo-spectral results.}
    \label{fig:skewness}
\end{figure}

To conclude, Fig.\ref{fig:skewness} shows a comparison between the z-distributions of the skewness of the components of the fluctuating velocity as predicted by the present LB model and the pseudo-spectral simulations of \cite{kim1987turbulence}. Once again, higher-order statistics are captured with a good degree of accuracy, as clearly visible in the figure. We argue that the small deviations from the reference data could be attributed to the different domain sizes employed in this work (both in the streamwise and crossflow directions) and the sampling procedure,  which may affect the averages. 

\subsection{\label{sec:jet} Axisymmetric turbulent jet at $Re=7000$ }

The case considered in this section is the space-time evolution of a  round turbulent jet featuring an inlet Reynolds number $Re_{jet}=\frac{U_{jet}D}{\nu}=7000$, being $U_{jet}=0.05$ (inlet velocity), $D=20$ (diameter of the round nozzle) and $\nu=1.3\cdot10^{-4}$ corresponding to a relaxation time $\tau=0.5004$ ($\nu=1.3\cdot10^{-4}\;lu^2/step$). Considering these numbers, the convective Mach number results in $Ma=U_{jet}/c_s\sim 0.09$, well below the threshold value $0.6$, which triggers non-negligible effects due to compressibility. Thus, comparing the present simulations against incompressible jet experiments reported in \cite{hussein1994velocity} and \cite{panchapakesan1993turbulence} is reasonable. The simulation has been performed on a Cartesian (uniform) grid ($N_x\times N_y\times N_z$) of $440\times440\times 600$ computational nodes.
Inflow boundary condition has been imposed at the inlet of the jet $z=0$ via the thread-safe non-equilibrium 
boundary conditions described in section \ref{sec:threadsafeBCs}, while periodic boundary conditions are enforced along both crossflow directions (i.e., along $x$ and $y$). 
We also impose an outflow condition at $z=N_z$ by 
extrapolating the linear momentum and the values of the components of the momentum flux tensor from neighboring nodes adjacent to the boundary nodes. Moreover, to reduce the effect of back-propagating pressure waves from the outlet, we enforced a sponge layer by increasing the viscosity near the exit as follows:

\begin{equation}
    \nu_{sponge}=\nu \left(\mathcal{K} \left(\frac{z-z_{start}}{N_z-z_{end}}\right)^p +1 \right),
\end{equation}

being $\mathcal{K}=1000$, $p=-3$ and $z_{start}=N_z-50$ grid points \cite{xue2022synthetic}.

\begin{figure}
    \centering
    \includegraphics[scale=0.72]{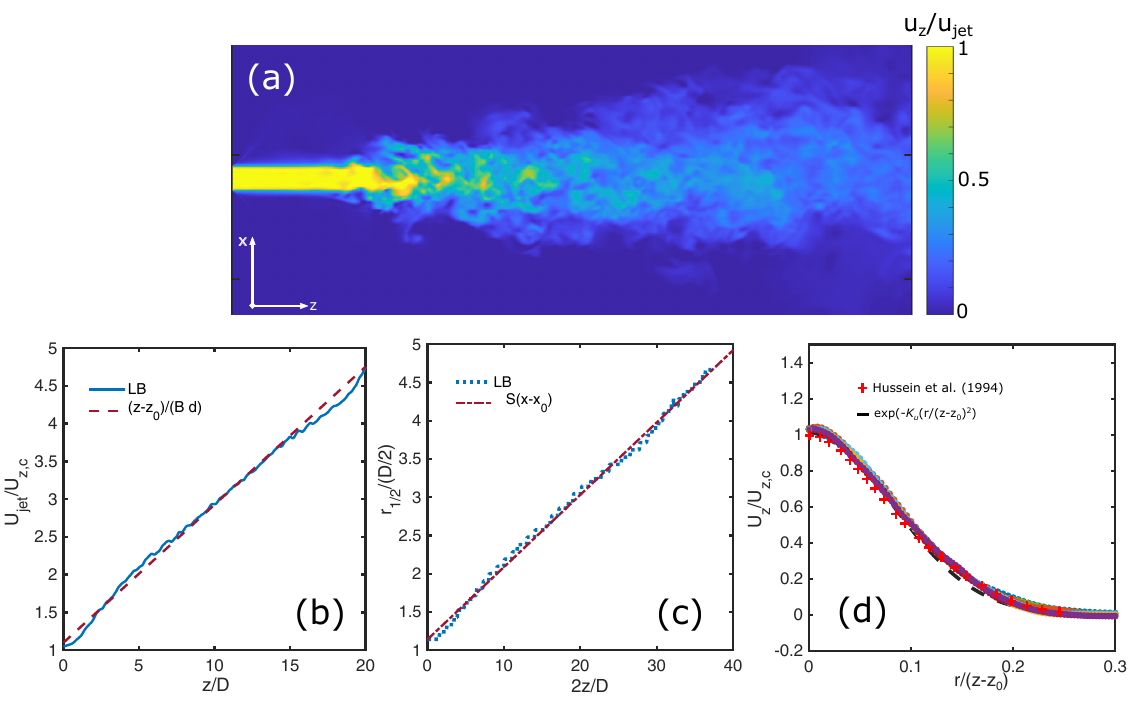}
    \caption{Panel (a) shows the instantaneous velocity field (streamwise component) of the turbulent jet at $Re=7000$. Panel (b) and (c) show the inverse of centerline velocity (made non-dimensional by the inlet velocity) and the spreading radius of the jet vs the distance from the inlet. Both plot reports the straight line behaviors occurring in the fully developed region of the jet \cite{pope2000turbulent}. Panel (d) reports the mean axial velocity profiles at several locations downstream of the nozzle. The profiles fall onto the same curve, denoting the self-similarity of the mean axial velocity in turbulent incompressible jets \cite{pope2000turbulent}.}
    \label{fig:wavg}
\end{figure}

Finally, for the inlet velocity we  set a uniform profile
with a superimposed, time-varying Gaussian noise \cite{wang2010direct}, needed to trigger the onset of instability at the outlet and aimed at mimicking the fully turbulent flow inside the round inlet pipe (not simulated to reduce the computational burden).

The results of the simulation, such as mean velocity and stresses, are compared with experimental data of Hussein et al. \cite{hussein1994velocity} and of Panchapakesan and Lumley \cite{panchapakesan1993turbulence}.

\subsubsection{Mean velocity, turbulent kinetic energy and Reynolds stresses}

Fig.\ref{fig:wavg}b shows the centerline velocity as a function of the streamwise distance, taken from $z_0\sim 5 D$, which is the end of the developing region \cite{pope2000turbulent}. 
The mean axial velocity obeys a linear relation of the kind

\begin{equation}
    U_{jet}/U_{z,c}= \frac{z-z_0}{B D}, 
\end{equation}

where $B$ is a constant measuring the 
decay rate $1/B$ and $z_0$ is the location of the virtual origin. The value of $B$ 
is $\sim 5.5$, in agreement with the ones reported in \cite{hussein1994velocity} ($B=5.8$), \cite{wygnanski1969some}($B=5.4$) and \cite{panchapakesan1993turbulence} ($B=6$). 

In Fig.\ref{fig:wavg}c,  we show the variation of the jet radius $r_{1/2}$, i.e. the radial distance from the axis where the mean axial velocity is equal to half the value assumed on the axis of the jet, as a function of the distance from the inlet (starting from the end of the developing region). As expected, the jet spreads linearly with a characteristic rate $S=0.094$, in excellent agreement with experimental predictions \cite{pope2000turbulent, hussein1994velocity}.
\begin{figure}
    \centering
    \includegraphics[scale=0.85]{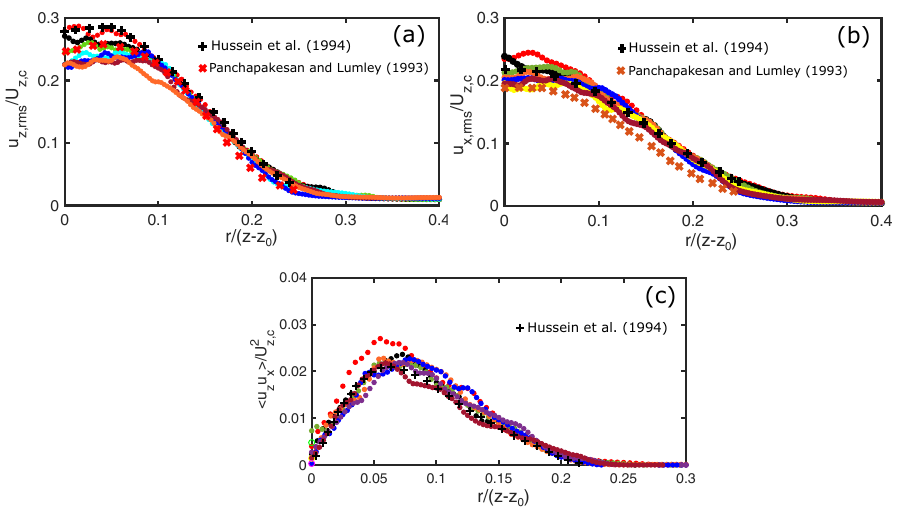}
    \caption{Panel (a) and (b) show the axial and radial rms values of the fluctuating velocity components, while panel (c) reports the Reynolds stresses (all made non-dimensional by the centerline axial velocity), taken at different sections along the jet axis. The LB results (filled circles) are compared against the experiments of Hussein et al. \cite{hussein1994velocity} and Panchapakesan and Lumley \cite{panchapakesan1993turbulence}}
    \label{fig:stressjet}
\end{figure}
Panel (d) of figure \ref{fig:wavg} displays the mean axial velocity profiles at several locations downstream of the nozzle. As one can see, all profiles fall onto the same curve, denoting the self-similarity of the mean axial velocity profile in turbulent incompressible jets. The velocity profiles are compared against experimental data reported in \cite{hussein1994velocity} (plusses symbols) and an exponential fit $U_{z}/U_{z,c}=exp\left(-k_u \left(\frac{r}{z-z_0}\right)^2 \right)$. The best fit is obtained with $k_u \sim 75$, in close agreement with the values reported in \cite{panchapakesan1993turbulence, boersma1999large}. 

The above scaling laws and self-similar behavior of the mean velocity profile represent a zeroth order test to benchmark against the thread-safe LB. Below, we proceed to compare the numerical prediction of the turbulent stress components against available experimental data \cite{hussein1994velocity, panchapakesan1993turbulence}.

In Fig.\ref{fig:stressjet}, plots of the turbulent kinetic energy components (streamwise and radial) and the Reynolds stresses taken at several locations downstream of the nozzle after the developing region are provided. The second-order moments are made non-dimensional by the mean square centerline velocity $U_{z,c}^2$. As recalled above, the LB results have been superposed to the experimental data reported in \cite{hussein1994velocity} (black plusses) and \cite{panchapakesan1993turbulence} (red crosses). 

The present results agree very well with the experiments 
discussed in 
Hussein et al., as evidenced by the overlapping between simulation and experimental data sets. In particular, it is interesting to note that the LB simulations predict the position of the peak of the Reynolds stress around $r/(z-z_0)=0.06$ while they zeroed at  $r/(z-z_0) \sim 0.2$, values close to those observed in incompressible turbulent jets experiments (see \cite{pope2000turbulent}). 


\section{\label{sec:performances} A note on the implementation and the performances on GPU of the High-order thread-safe LB}

The codes have been run on two different accelerators: GeForce RTX 3090 and Nvidia A100 (on Leonardo supercomputer). The former features 24 Gbytes of RAM shared by 10496 Cuda cores with a peak performance $\sim 35$ TeraFLOPS in single precision and a memory bandwidth of $\sim 900$ GB/s, while the latter has 64 Gbytes of RAM shared by 6912 Cuda cores with a peak performance $\sim 19.5$ TeraFLOPS in single precision with a memory bandwidth of  $\sim 2.0 TB/s$. 

The performances of the GPU porting implementations have been measured in billions of lattice update per second (GLUPS), defined as:

\begin{equation}
    GLUPS=\frac{n_x n_y n_z n_{steps}}{10^9 T_{sim}},
\end{equation}

where $n_{x,y,z}$ are the number of lattice nodes along the three spatial dimensions, $n_{steps}$ is the number of simulation time steps and $T_{sim}$ is the run wall-clock time (in seconds) needed to perform the simulation.

The computational performances have been analyzed by running the pressure-driven flow benchmark on a domain having $Nx\times Ny \times Nz=300\times 10^6$ grid points.

The code has been ported on GPU by exploiting OpenACC directives that, as shown below, allow peak performances 
in line with the state-of-the-art LB computing with virtually no extra cost in terms of coding.
Implementation details can be found in \cite{MONTESSORIjocs23}, and the actual formulas for the high-order equilibria and non-equilibrium distributions are reported in the appendix. Moreover, the interested reader is referred to the publicly available code shared on Git-Hub (see code availability section).

In the case of the pressure-driven flow, we reached a peak performance of $\sim 2$ GLUPS on RTX3090 and $\sim 3$ on the A100 mounted on Leonardo. We then inspected the roofline model of the stream-collide kernels, which exposes an operational intensity $\sim 7 FLOP/byte$ achieving almost the ideal performance (bounded by the memory-bandwidth roof of the plot) of $5 \; Teraflop/s$, booth larger than those reported in \cite{MONTESSORIjocs23} ($\sim 1.5 flop/byte$ and $\sim 1.5 Teraflops/s$) and of other state-of-the-art lattice Boltzmann HPC codes \cite{lehmann2021ejection,lehmann2022esoteric,latt2021palabos}. 
Thus, the larger number of operations needed to compute the high order equilibria and non-equilibrium distributions results in an increase of the number of floating point operations per unit time while the transferred amount information remains unchanged. In turn, the arithmetic intensity of the algorithm increases as well allowing to approach to the ideal peak performance delivered by the device.
Lastly, it is worth noting that while state-of-the-art implementations are usually based on standard BGK collisional operator, our approach exploits a high-order regularized procedure aimed at increasing the stability and accuracy of the LB, thus allowing to span a broader and lower range of viscosities, as low as the ones reachable via MRT-like models, along with a sensible reduction of the memory footprint with respect to the standard LB model based on the AB streaming strategy.

\section{Conclusions}

This work presented a high-order, thread-safe implementation of the lattice Boltzmann model for HPC simulations of turbulent flows, with the 
aim of exploiting in full the memory bandwidth of GPU-based architectures.
Our results show 
an outstanding performance while retaining the coding simplicity of standard lattice Boltzmann implementations, with a sensible reduction of the memory footprint if compared to fused streaming-collision strategies and with no need to resort to more complex and exotic streaming algorithms \cite{geier2017esoteric,lehmann2022esoteric}. 
Moreover, a new boundary condition based on the concept of non-equilibrium extrapolation method and Grad's reconstruction of missing data\cite{chikatamarla2006grad} has been developed, that allows for an efficient implementation of Dirichlet and boundary conditions by imposing conditions at the boundaries directly on the macropscopic fields of interest.
The above HPC implementation has been benchmarked against two different case studies, namely the turbulent flow in a straight channel at $Re_{\tau}=180$ and the axisymmetric turbulent jet at $Re=7000$. The results confirm the accuracy of the high-order TSLB and, in turn, its ability to predict the behavior of highly turbulent flows. Future work will be focused on the possibility of scaling the single-GPU performances of the thread-safe LB on multi-GPU architectures with the aim of scaling up to very large-scale simulations of turbulent flows calling for the computational power of existing pre-exascale supercomputers and upcoming exascale infrastructures.

\section*{Acknowledgments}

A.M. and M.L. acknowledges the CINECA Computational Grant IsCb3 "recTSLB", IsCb4 "LLBfast" and IsCa9 "3DMPILB" under the ISCRA initiative, for the availability of high performance computing resources and support.
S.S. acknowledges funding from ERC-PoC2 grant No. 101081171 (DropTrack). 

\section*{Code availability}
The code is freely available at \url{https://github.com/andreamontessori/entropicThreadsafe}

\bibliographystyle{apsrev4-1}
\bibliography{biblio.bib}

\end{document}